\documentclass[aps,twocolumn,preprintnumbers,prd,nofootinbib,10pt]{revtex4-2}

\usepackage{aas_macros}

\usepackage{graphicx,epsfig}
\usepackage{amsmath,amssymb}
\usepackage{amsfonts}
\usepackage{float}
\usepackage{fancyhdr}
\usepackage{xcolor}
\usepackage{lineno}
\usepackage{subfigure}
\usepackage[english]{babel}
\usepackage[normalem]{ulem} 
\usepackage{bm}

\usepackage{hyperref}

\hypersetup{
 colorlinks=true,
 linktoc=all,
 linkcolor=red,
 citecolor=blue,
 urlcolor = blue}

\newcommand{\ii}{\mathrm{i}} 
\newcommand{\dd}{\mathrm{d}} 

\newcommand{\bfxi}{\mbox{\boldmath{$\xi$}}}
\newcommand{\bfeta}{\mbox{\boldmath{$\eta$}}}

\newcommand{\beq}{\begin{equation}}
\newcommand{\beqa}{\begin{eqnarray}}
\newcommand{\eeq}{\end{equation}}
\newcommand{\eeqa}{\end{eqnarray}}

\begin{document}

\title{Probing Binary Lens Caustics with Gravitational Waves: A Uniform Approximation Approach}

\author{Anna Moreso Serra}
\email{amoreso@icc.ub.edu} 
\author{Oleg Bulashenko}
\email{oleg@fqa.ub.edu} 
\affiliation{Institut de Ci\`encies del Cosmos (ICCUB), Facultat de F\'{\i}sica, Universitat de Barcelona, Mart\'i i Franqu\`es 1, E-08028, 08028 Barcelona, Spain.}

\date{\today}

\begin{abstract}
\section*{Abstract}
We present a new framework for modeling gravitational wave diffraction near fold caustics using the Uniform Approximation (UA), focusing on binary mass lenses—axially asymmetric systems with complex caustic structures. Full-wave methods based on the Kirchhoff integral become impractical in this regime due to highly oscillatory integrands. The UA provides a robust and accurate description of the wave field near folds, resolving the breakdown of Geometrical Optics at caustics and improving upon Transitional Asymptotics—based on Airy function approximations—which lack global validity.
Central to our approach is the concept of the caustic width, $d_c$, a characteristic length scale defining the region where diffraction significantly alters wave propagation. We find that $d_c$ scales universally with the gravitational wavelength as $\sim \lambda^{2/3}$ and inversely with the redshifted lens mass as $\sim M_{Lz}^{-2/3}$.
The wave amplification near the fold grows as $\sim d_c^{-1/4}$, substantially enhancing the signal and potentially playing a key role in the detection of gravitational waves lensed near caustics. Notably, for lens masses below the galactic scale, the caustic width for gravitational waves is not negligible compared to the Einstein radius—as it is in electromagnetic lensing—making the UA essential for accurately capturing wave effects.

\end{abstract}


\maketitle



\section{Introduction}
\label{sec:introduction}

As gravitational waves (GWs) propagate through space, they may encounter massive objects that significantly alter their properties, much like electromagnetic (EM) waves. The phenomenon of gravitational lensing has been extensively studied and successfully applied to EM waves, yielding groundbreaking results in astronomy and astrophysics. One of its more powerful applications is in cosmography, where time delays between multiple images of lensed quasars and supernovae enable precise measurements of the Hubble constant \cite{refsdal_64,suyu_2010,Pascale_2025,Birrer2024} and place constraints on other cosmological parameters \cite{grillo_2024}. 
Another use of strong lensing by galaxies and galaxy clusters is to map the distribution of dark matter across different scales \cite{schneider_mao_98,dalal_2002,Vegetti2024}. 
It has also been employed to investigate the mass structure and evolution of elliptical galaxies, constrain the stellar initial mass function (IMF) \cite{Shajib2024}, and study their interstellar medium \cite{Falco_99}. 

One particularly fascinating aspect of gravitational lensing is the substantial magnification observed near caustics. This effect has been thoroughly explored in the context of EM waves. Due to the extraordinarily high magnification—often exceeding $10^4$ times—objects such as individual stars and distant galaxies can be detected at redshifts far beyond the reach of current detector sensitivity for unlensed observations \cite{Weisenbach2024}. 
This exceptional capability has opened up new possibilities for observing faint and distant astrophysical phenomena, allowing constraints on source size and structure while providing insights into the accretion disks and broad-line regions of quasars \cite{mortonson_2005,Suyu2024,Vernardos2024}.

Recent advancements in GW interferometry have brought these detectors to the level of sensitivity required to observe gravitational lensing effects in GWs. 
During the first three observing runs, the global network of GW detectors---Advanced LIGO~\cite{Aasi:2014jea}, Advanced Virgo~\cite{Acernese:2014hva}, and KAGRA~\cite{KAGRA:2020tym}---achieved a steadily increasing detection rate of compact binary coalescences, resulting in the observation of 90 events by the end of O3~\cite{GWTC1, GWTC2, GWTC2.1, GWTC3}. 
This significant increase in detection rate has markedly enhanced the probability of observing gravitational lensing signatures in GW signals \cite{lensingO3b,smith_rates-2023}.
A range of phenomena, from cosmological to astrophysical scales, could become accessible through the lensing of GWs.
These include, for instance, the measurement of galaxy velocity dispersions, constraints on dark matter and binary formation channels, the detection of intermediate-mass black holes, and novel tests of general relativity, among others (see, e.g., the references listed in~\cite{lensingO3b,smith_rates-2023}).

We focus here on the case of two point masses acting as a lens, commonly referred to as a binary mass system. This choice is motivated by several factors. Approximately half of solar-type stars are members of multiple systems, making binaries a frequent occurrence in stellar populations. Furthermore, the likelihood of forming multiple systems increases with stellar mass \citep{duchene2013stellar}.
Binary black holes can also serve as gravitational lenses. However, the prevalence of black holes in binaries depends on their formation channels and the environments in which they form—an area of active research \citep{2020FrASS...7...38M}. Notably, even two masses that are not gravitationally bound can function as a binary lens, emphasizing the flexibility and broad relevance of this configuration in gravitational lensing studies.

It is worth noting that the binary mass lens (BML) has received less attention in the literature compared to axially symmetric models, such as the point mass lens 
or the singular isothermal sphere. 
Unlike axially symmetric lenses, where the resulting caustics are relatively simple—typically forming circles or points—the breaking of axial symmetry in this model introduces a greater degree of topological complexity. This asymmetry gives rise to caustics with intricate and diverse structures, which depend sensitively on the specific configuration of the lens system.
In the case of a BML—the simplest example of an axially asymmetric lens—the topology of the caustics is governed by a single parameter: the separation between the two masses, expressed in units of the Einstein radius. This straightforward dependence makes the binary system an ideal framework for studying the effects of asymmetry on gravitational lensing phenomena while retaining mathematical and computational tractability.

When a GW source approaches a caustic, the signal detected by the interferometer is expected to show significant amplification. 
This phenomenon is well-studied in the field of EM waves but remains largely unexplored in the context of GWs. It is intriguing from an astrophysical perspective and is currently a subject of significant interest in ongoing research \cite{lo-rico24,ezquiaga25}.
However, near the caustic, the geometrical optics (GO) approximation no longer provides an accurate description
\cite{nakamura99, ezquiaga-holz-21,berry21,choi21}.
Specifically, as the source approaches a caustic point, the time delay between the merging (or coalescing) images becomes vanishingly small. This implies that achieving the GO limit would require infinitely high GW frequencies. 
Full-wave modeling provides greater accuracy than the GO approximation. However, it becomes computationally intensive and demands significant computational resources, particularly for complex mass distributions.
These limitations underscore the need for alternative strategies. Asymptotic methods, which bridge the gap between GO and full-wave approaches, offer a promising solution by capturing the essential physics without requiring exhaustive computations. 

One crucial difference between EM waves and GWs lies in the scale of their wavelengths. For EM waves, the wavelength is typically much smaller, so the region near the caustic where the GO approximation breaks down is very small, often negligible, especially when considering galaxy-scale masses. In this regime, where wavelengths are very short and masses are high, the geometrical optics approximation remains sufficiently accurate for analyzing the wave field down to very close distances from the caustic.
For GWs, however, the wavelength is much larger, typically $\gtrsim 10^6$ meters in the LIGO-Virgo-KAGRA (LVK) frequency band \cite{Abbott2020}. As a result, the region near the caustic where the geometrical optics approximation breaks down is significantly broader and may be substantial compared to the characteristic dimension of the lens. In this context, the GO approximation fails to provide an accurate description, and a wave-optics approach is required to capture the behavior of the waves near the caustic.

As an alternative to the GO approximation, one of the most notable methods is the {\em Transitional Approximation} \cite{berry21}, which describes the wave field near a caustic in terms of the Airy function \cite{airy_1838}. This solution provides a continuous representation across the fold, avoiding the divergences at the caustic seen in GO. 
The key difference lies in the expansion of the phase of the wave field as a function of the spatial coordinate near the caustic: the standard GO approximation expands the phase only up to the quadratic (second-order) term, which causes divergences when the second derivative vanishes at the caustic. In contrast, the Transitional Approximation includes the cubic (third-order) term in the phase expansion, which regularizes the solution and accurately captures the wave behavior near the fold.
The approximation using the Airy function is widely employed in the literature 
\cite{schneider-92,lo-rico24,ezquiaga25}. However, while it provides an excellent {\em local} solution in the immediate vicinity of the singular region near the fold singularity, it is not valid at larger distances and does not align with the GO approximation far from the fold, where the latter is certainly valid. The two solutions converge only in the limit of infinitely small wavelengths (a good approximation for EM waves) or high lens masses, comparable to that of a galaxy \cite{ezquiaga25}.
To fully characterize this phenomenon and provide a comprehensive description valid across a wide range of parameters, we introduce the {\em caustic width}, $d_c$, as a physical parameter that enables the study and analysis of various limiting cases.

In this paper, we introduce the Uniform Approxima\-tion: an alternative asymptotic method that accurately models the wave field near a fold caustic (also known as a fold catastrophe in the context of catastrophe theory \cite{berry80iv}) and smoothly connects with the GO solution far from the caustic. This approach offers a consistent and versatile solution, particularly suited for integration into GW lensing pipelines. Importantly, it relies on analytical functions, specifically the Airy function and its derivative, and avoids extensive computations. This makes it particularly advantageous for scenarios where the source crosses the fold caustic.

\section{Microlensing on BML}
\label{sec:microlens}

\begin{figure}
    \centering
    \includegraphics[width=\columnwidth]{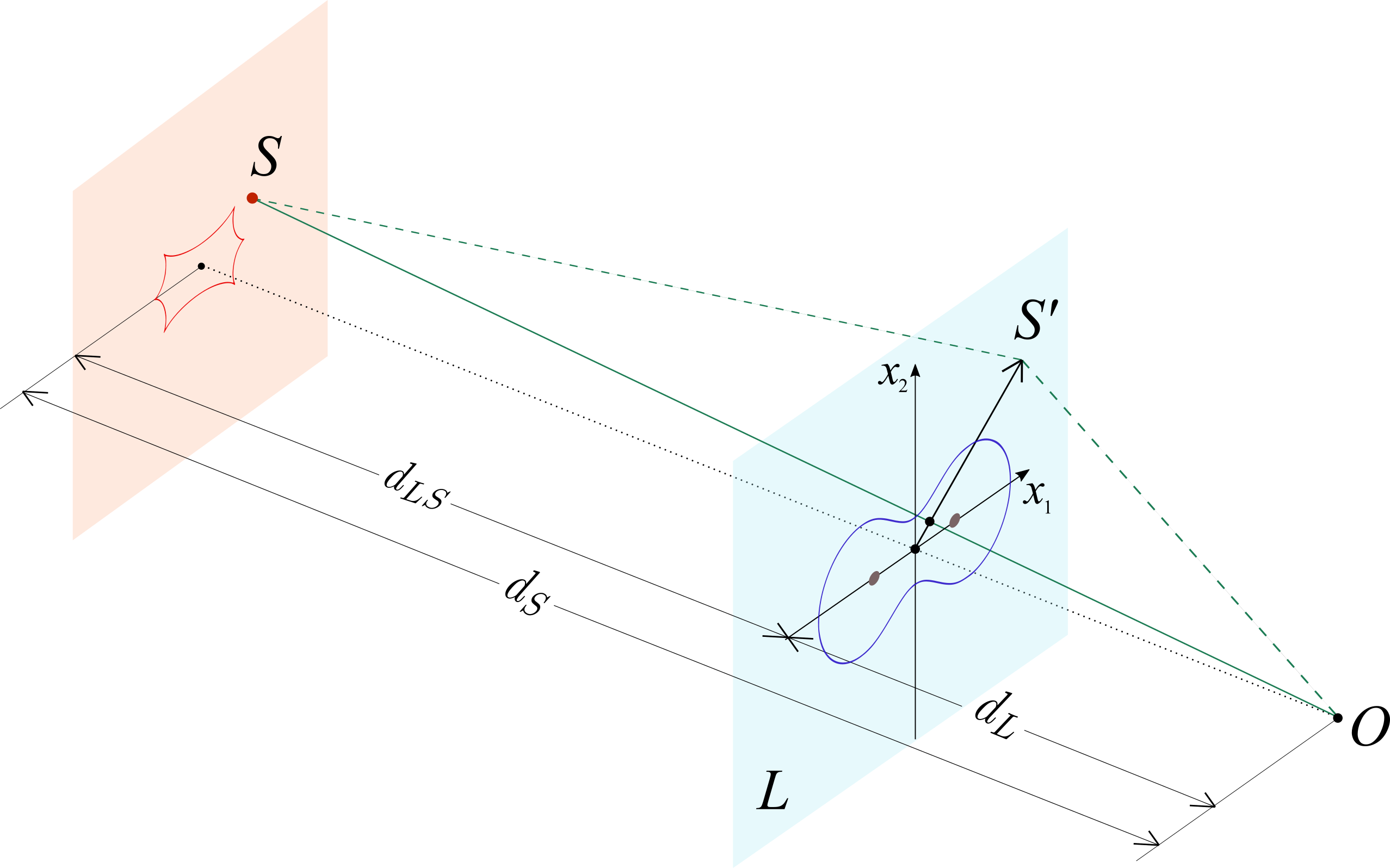}
    \caption{
    A schematic diagram of gravitational lensing in the thin-lens approximation. The dotted line indicates the line of sight, perpendicular to both the lens and source planes. In the absence of a lens, the wave emitted by the source $S$ would follow a single path, shown as the solid green line. Due to lensing, the total wave is the sum of all partial waves passing through the lens plane $L$. One such partial wave is depicted, deflected at point $S'$ in the lens plane (dashed green line), before reaching the observer $O$. 
    }
    \label{fig1:geom}
\end{figure}

A binary mass system with sufficient mass can act as a lens for a passing gravitational wave.
Lensing effects occur when the source, the lens and the observer are all aligned within the Einstein angle $\theta_{\rm E}=R_{\rm E}/d_L$, i.e., the lens is located near the line of sight. 
The Einstein radius $R_{\rm E}$ is commonly expressed as:
\begin{equation}
    R_{\rm E} = \sqrt{\frac{4GM_L}{c^2}\, \frac{d_{LS}\,d_L}{d_S} },
\label{R_Einst}
\end{equation}
where $M_L$ is the mass of the lens, $d_{LS}$ is the angular diameter distance between the source and the lens, and $d_L$ and $d_S$ are the angular diameter distances to the lens and source at redshifts $z_L$ and $z_S$, respectively. 
$R_{\rm E}$ represents the characteristic length scale on the lens plane, is proportional to $\sqrt{M_L}$ and is typically much smaller than the cosmological distances $d_{LS}$, $d_L$, and $d_S$.  
This enables the lens mass to be projected onto a lens plane. In the thin lens approximation, gravitational waves propagate freely outside the lens, interacting only with a two-dimensional gravitational potential at the lens plane, where the lensing effect is ultimately captured in the transmission factor $F$ \cite{schneider-92}.
Another important length scale is the distance between the binary masses, which will be introduced below. For a schematic diagram of lensing by a binary mass lens, we refer the reader to Fig.~\ref{fig1:geom}.

\subsection{Wave approach to gravitational lensing}
An unlensed GW signal from the source can be described by its frequency-domain strain $\tilde{h}(f)$, which is the Fourier transform of the time-domain strain $h(t)$.
After the signal passes through a lens, the resulting lensed waveform $\tilde{h}_L(f)$, which is ultimately detected, is the product of the transmission factor $F(f)$ and the original unlensed waveform:
\beq
    \tilde{h}_L(f) = F(f) \cdot \tilde{h}(f).
\label{lensed_strain}
\eeq
The transmission factor, also known as the amplification factor \citep{takahashi03}, is defined as the ratio of the lensed to the unlensed gravitational wave amplitudes at the observation point. It is determined by the Fresnel-Kirchhoff diffraction integral 
\footnote{In our code we take the complex conjugate by replacing $\ii$ with $-\ii$ to align with the Fourier transform convention used in the Python libraries of the LVK Collaboration~\cite{lalsuite-fourier}.}
across the lens plane \cite{born-wolf-03,schneider-92}
\beq
F(f,\mathbf{y}) = -\ii f\, t_M
\iint e^{\ii \,2\pi f\, t_d (\mathbf{x},\mathbf{y})} \, \dd^2\mathbf{x},
\label{eq:F_phys_units}
\eeq
where the lensing time delay function is given by
\beq
t_d(\mathbf{x},\mathbf{y}) = t_M \, \left( \frac{1}{2}\, | \mathbf{x}-\mathbf{y} |^2
- \psi(\mathbf{x}) + \phi_m(\mathbf{y}) \right)
\label{t-delay}
\eeq
with the characteristic lensing time in physical units
\beq
t_M = (1+z_L) \,\frac{\xi_0^2\,d_S}{c\,d_L d_{LS}}.
\label{tM_xi}
\eeq
Here, $\xi_0$ is an arbitrary length scale in the lens plane, used to normalize both the impact parameter $\bfxi$ and the source position $\bfeta$ as follows \cite{takahashi03}:
$\mathbf{x}=\bfxi/\xi_0$, $\mathbf{y} = \bfeta \,d_L/(\xi_0 d_S)$.
In this notation, $\bfeta$ indicates the position of the source in the source plane, while $\mathbf{y}$ represents the projection of the source onto the lens plane, normalized by $\xi_0$.
The factor $(1+z_L)$ is included to account for cosmological distance, where $z_L$ represents the redshift of the lens.
The properties of the lens are captured by the lensing potential $\psi(\mathbf{x})$. 
For convenience, one may define the phase $\phi_m(\mathbf{y})$ such that the minimum of $t_d(\mathbf{x}, \mathbf{y})$ at fixed $\mathbf{y}$ is zero.

Next, we will consider a binary mass system as the lens. 
We can treat each lensing mass as point-like when its physical radius is much smaller than its respective Einstein radius, as in the case of binary black holes, stars, dense dark matter clumps, and similar compact objects. 
The single point mass lens (PML) model has been widely used in the literature to interpret both electromagnetic \cite{deguchi86a,deguchi86b,schneider-92,petters-01} and GW lensing 
\cite{nakamura98,nakamura99,takahashi03,
matsunaga06,christian18,diego19,cheung20,cremonese21a,seo21,wright21,yeung21,BU-JCAP-21,caliscan22,tambalo22a,bondarescu22,savastano23,mishra23a,chen24,villarrubia24,chan-seo-25,bada25}. 
However, the binary mass lens (BML) model has been less commonly applied in comparison \cite{Kim24,meena25}.

For the BML model a natural choice for the length scale $\xi_0$ is the Einstein radius given by Eq.~\eqref{R_Einst},
where the mass is the sum of the two point masses forming the lens, $M_L=M_1+M_2$. 
In the lens plane, we define a coordinate system $(x_1, x_2)$ such that the $x_1$ axis is aligned with the line connecting the two masses, and the origin is placed at the midpoint of this line.
Under these assumptions, the lensing potential in Eq.~\eqref{t-delay} can be expressed as
\cite{schneider-weiss-86,mollerach2002book}
\begin{equation}
    \psi(\mathbf{x})=\gamma_1\ln\left|\mathbf{x}-\mathbf{x_L}\right|+\gamma_2\ln\left|\mathbf{x}+\mathbf{x_L}\right|,
    \label{lensing_potential}
\end{equation}
where $\gamma_1=M_1/M_L$, $\gamma_2=M_2/M_L$, and $\mathbf{x_L}=(b,0)$.
With this normalization, the physical distance $d$ between the masses is given by $d = 2b R_{\rm E}$, where $b$ is a dimensionless parameter \cite{schneider-weiss-86}.
By substituting $\xi_0 = R_{\rm E}$ into Eq.~\eqref{tM_xi}, the time $t_M$ becomes directly proportional to the lens mass:
\begin{equation}
t_M =2 R_{\rm S}/c \;\approx \; 1.97 \times 10^{-5} \,{\rm s}\; (M_{Lz}/M_\odot),
\label{tM}
\end{equation}
where $R_{\rm S} =2GM_{Lz}/c^2$ is the Schwarzschild radius, and $M_{Lz} = M_L (1+z_L)$ represents the redshifted mass of the lens.
The diffraction integral \eqref{eq:F_phys_units}, expressed in terms of dimensionless quantities, is given by:
\begin{equation}
    F(w,\mathbf{y})=\frac{w}{2\pi \ii}\iint e^{\ii w T(\mathbf{x},\mathbf{y})}\, \dd^2\mathbf{x}
    \label{eq:F-adim}
\end{equation}
where $w\equiv 2\pi f\,t_M$ is the frequency of the gravitational wave scaled by the characteristic time $t_M$, and the dimensionless time delay for the symmetric BML ($\gamma_1=\gamma_2=1/2$) takes the form
\begin{align}
    T(\mathbf{x},\mathbf{y}) = &
    \frac{1}{2}\left[(x_1-y_1)^2+(x_2-y_2)^2\right] \notag\\ &
    -\frac{1}{4}\ln{\left[(x_1^2+x_2^2+b^2)^2-4x_1^2b^2\right]} .
    \label{eq:t_pre_expansion}
\end{align}
This function represents the travel time of a GW ray passing through the lens plane at $\mathbf{x}=(x_1,x_2)$, relative to a ray crossing the lens plane at $\mathbf{x}=\mathbf{y}=(y_1,y_2)$, which is unaffected by the lens's gravitational potential \cite{schneider-92}.
As seen from Eq.~\eqref{eq:F-adim}, 
the transmission factor, which is a function of frequency $f$, depends on the parameters:
(i) the mass of the lens $M_{Lz}$ through the time $t_M$, (ii) the scaled position of the source $\mathbf{y}$, and (iii) the dimensionless distance between the masses, $2b$.

We note that, in our analysis, the lens configuration can be considered static during the wave passage because the orbital period of the lens binaries is much longer than the timescale associated with the motion of the source binaries (i.e., the GW period and chirp duration).  This difference in timescales reflects the vastly different length scales involved: the lens binary separation is on the order of $R_{\rm E}$, 
while that of the source is on the order of $R_{\rm S}$.

\subsection{Geometrical Optics Approximation}
In principle, Eq.~\eqref{eq:F-adim} provides a means to calculate the full-wave transmission factor for arbitrary mass distribution in the lens plane. However, in practice, when the time delays between different paths induced by the gravitational potential become large compared to the wave period, the integral involves a rapidly oscillating function. This significantly increases the computational cost, as the integral converges very slowly.
Fortunately, in this limit, the dominant contribution comes from a few well-defined images of the source, which correspond to the stationary points of the time delay function,
$\partial T(\mathbf{x},\mathbf{y}) / \partial \mathbf{x} =0$ at $\mathbf{x}=\mathbf{x}_j$ for a given $\mathbf{y}$.
The transmission factor in this GO limit
can then be written as a sum over these stationary points \cite{schneider-92,nakamura99},
\begin{equation}
F_{\rm GO}=\sum_j |\mu_j|^{1/2} \exp{\left[\ii w \,T (\mathbf{x}_j,\mathbf{y})- \ii \pi n_j/2 \right]},
\label{F_GO}
\end{equation}
each characterized by a different magnification $\mu_j$ and a Morse index $n_j$, which is associated with a topological phase shift.
The sum in Eq.~\eqref{F_GO} is taken over the stationary points, whose positions $\mathbf{x}_j$ are determined as the real roots of the lens equation (see Appendix \ref{appendixA} for details).
The magnification of the $j$-th image is obtained from the determinant of the Hessian of the time-delay function evaluated at the stationary points
\beq
    \mu_j=[\det \Hat{K}(\mathbf{x}_j)]^{-1},
\label{eq:magnif}
\eeq
where the Hessian matrix is defined as: 
\beq
\Hat{K} = 
\begin{bmatrix}
T_{11} & T_{12} \\
T_{21} & T_{22}
\end{bmatrix}
\label{eq:hessian}
\eeq
with its components given by
$T_{ab} \equiv \partial^2 T/\partial x_a \partial x_b$.
The Morse index is equal to the number of negative eigenvalues $\Lambda_i$ of the matrix $\Hat{K}$. It can be formally calculated as:
$n_j = 2 -{\cal H}(\Lambda_1)-{\cal H}(\Lambda_2)$, where ${\cal H}$
denotes the Heaviside step function. This results in $n_j=0,1$ or $2$, corresponding to type I, II, and III images, which are a local minimum, saddle point, and maximum of $T(\mathbf{x},\mathbf{y})$ in the $(x_1,x_2)$ plane, respectively.

For certain specific positions of the source relative to the lens, $\det \Hat{K}$ vanishes. 
The set of these points in the source plane forms the {\em caustics}. The projections of the caustics onto the lens plane are referred to as the {\em critical lines}. When the source crosses the caustic, the number of images changes by two \cite{schneider-92}. Specifically, along the critical line, two of the images coalesce, resulting in  $\det \Hat{K}=0$,  which causes the magnification to formally become infinite, as given by Eq.~\eqref{eq:magnif}. This indicates that the GO approximation is no longer valid for these two points.
The structure of caustics for a binary lens is explained in Appendix \ref{appendixA}. 
According to catastrophe theory, in the context of gravitational lensing, where there are only two control parameters related to the position of the source, 
the stable caustic singularities are classified into two types: folds and cusps \cite{berry80iv}. The smooth segments of the caustic curve are referred to as folds, whereas the points where two adjacent folds converge are known as cusps. Both types of singularities exhibit generic and universal properties, which is desirable as it enhances predictability, simplifies analysis, and facilitates comparative studies. 

The GO formula \eqref{F_GO} greatly simplifies and accelerates computations by reducing the task to summing a few terms, rather than performing an integration over the entire plane. It is particularly effective for obtaining the overall diffraction pattern at high frequencies and far from caustics (the criterion for the validity will be defined later).

\begin{figure*}[t]
    \centering
    \includegraphics[width=0.9\textwidth]{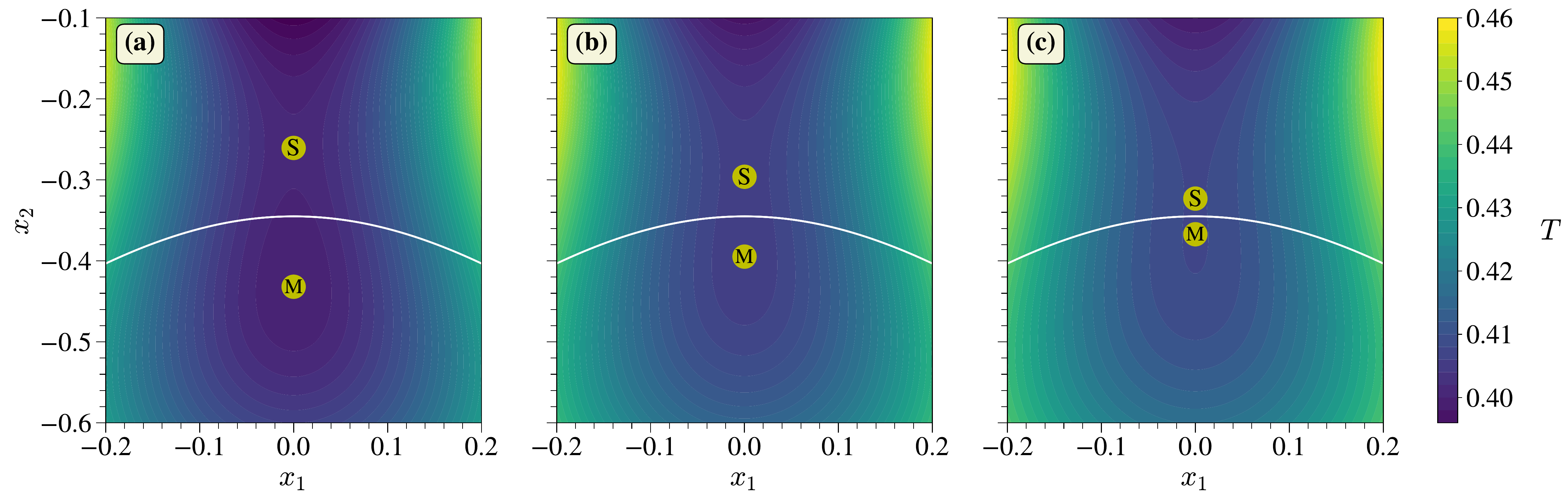}
    \caption{Time delay contours in the lens plane near the critical line (white) for a BML with a mass separation of $b = 0.7$. The time delay values are computed using Eq.~\eqref{eq:t_pre_expansion}. The yellow circles indicate the two stationary points, which move closer together as the source approaches the fold:
    (a) $\tilde{y}_2 = -1.5 \times 10^{-2}$, (b) $\tilde{y}_2 = -5 \times 10^{-3}$, and (c) $\tilde{y}_2 = -10^{-3}$. The letters \textit{S} and \textit{M} denote the saddle and minimum nature of each image, respectively.}
    \label{fig:time_delay_3}
\end{figure*}

\section{Transitional Asymptotics}
\label{sec:TA}

The most intriguing scenario arises when the source is near the caustic, as it produces the highest signal amplification, occurring precisely where multiple rays converge.
We aim to describe the wave properties near caustics, where the GO approximation fails, using asymptotic methods rather than relying on the full-wave approach. 
Near a fold, two stationary points become closely spaced, making it impossible to separate their contributions using the simplest form of the stationary phase method, as applied in the GO limit. 
In the BML model under consideration, one of these points corresponds to a saddle point within the interior region, 
while the other represents a minimum outside it (see Fig.~\ref{fig:time_delay_3}). These points are denoted as $S$ and $M$, respectively, with the condition that the time delays satisfy $T_M < T_S$. 
As these points merge, the phase extremum in the Kirchhoff integral becomes degenerate.
This means that the first and second derivatives of the phase---or equivalently, the first- and second-order terms in its Taylor expansion around the stationary point---both vanish.
The evolution of this time delay is illustrated in  Fig.~\ref{fig:time-delay_x2} in Appendix \ref{appendixA}.
To accurately model this, we must approximate the phase using a polynomial expanded up to a non-zero third-order term. This higher-order approximation is crucial for preserving the stationary-point structure. The resulting integral, being more complex than a simple Gaussian integral, then requires evaluation using special functions.

To illustrate the method, we analyze a specific mass separation of $b = 0.7$ which satisfies $1/8\leq b^2\leq 1$, where the caustics form an astroid with six cusps connected by six folds, as shown in Fig.~\ref{fig:magnification_maps}(d).
For mathematical convenience, we consider the source moving along the symmetry line relative to the binary masses---$y_2$-axis---as it crosses the fold at $(0,y_2^0)$ (see Fig.~\ref{fig:images_app}). While this choice simplifies the analysis, we believe the approach can be generalized to other configurations by appropriately rotating the coordinate system. In this setup, the coalescence point of the two images in the lens plane is $(0,x_2^0)$. The coordinates of these points can be expressed in terms of the parameter $b$ as \cite{schneider-weiss-86}
\begin{equation}
\begin{gathered}
    x_{2}^0 = - \frac{1}{\sqrt{2}} \left[ 
    (1 + 8b^{2})^{1/2} -1 - 2b^{2} 
        \right]^{1/2},\\
y_2^0 = \frac{1}{2\sqrt{2}b} \left[ (1 + 8b^{2})^{3/2} + 1 - 20b^{2} - 8b^{4}\right]^{1/2}.
\end{gathered}
\label{eq:fold_coordinates}
\end{equation}
For $b=0.7$, we obtain the values: $x_2^0 \approx -0.34504$ and $y_2^0 \approx 0.22148$.

To get the Transitional Approximation \cite{berry80iv}, we expand the time delay function \eqref{eq:t_pre_expansion} as a Taylor series around $y_2^0$ in the source plane and $x_2^0$ in the lens plane. 
It is convenient to define 
the partial derivatives as
$T_{i\dots j} \equiv \partial_{x_j}\dots\partial_{x_i} T$ and to introduce the shifted coordinates $\tilde{y}_2=y_2-y_2^0$ and $\tilde{x}_2=x_2-x_2^0$.
Imposing the fold conditions $T_1 = T_2 = T_{12} = T_{22} = 0$, $T_{11} \neq 0$, $T_{222} \neq 0$, we obtain, to the lowest nontrivial order
\cite{schneider-92,Ulmer95,nakamura99,ezquiaga25}:

\beq
   T\approx T_0
   -\tilde{x}_2 \tilde{y}_2 
   +\frac{1}{2} \, T_{11} \,{x_1}^2
   +\frac{1}{6} \, T_{222} \,{\tilde{x}_2}^3.
\label{eq:fold_expansion}
\eeq
Here, $T_0$ represents a global time delay that is independent of the integration variables of the diffraction integral \eqref{eq:F-adim}. 
By substituting the expanded time delay \eqref{eq:fold_expansion} into Eq.~\eqref{eq:F-adim}, the original two-dimensional integral simplifies. The integration over $x_1$ 
results in a Gaussian integral, which can be evaluated analytically, leaving only an integral over $\tilde{x}_2$:
\beq
    F=\sqrt{\frac{w}{2\pi\ii \,|T_{11}|}}
    \int_{-\infty}^\infty e^{\ii w (T_0 -\tilde{x}_2 \tilde{y}_2 
    +T_{222} \,{\tilde{x}_2}^3/6)} \,
    \dd \tilde{x}_2.
    \label{eq:Gaussian_int}
\eeq
With the variable substitution, 
$\tilde{x}_2 = \tilde{t}\,(wT_{222}\,/2)^{-1/3}$,
Eq.~\eqref{eq:Gaussian_int} reduces to the canonical Airy integral \cite{airy_1838}, where
the Airy function 
\begin{equation*}
    \mathrm{Ai}(u)=
    \frac{1}{2\pi}\int_{-\infty}^\infty 
    \exp{[\ii\left(u \tilde{t} + \tilde{t}^3/3 \right)]} \,\dd \tilde{t}
\end{equation*}
provides an accurate description of wave behavior near the fold singularity. 
For $u<0$ (inside the caustic), it exhibits oscillations due to the interference of the two merging stationary points, while for $u>0$ (outside the caustic), it decays exponentially. At $u=0$, it smoothly transitions between these regimes, offering a {\em Transitional Approximation} (TA) across the fold \cite{berry21} (see e.g. Refs.~\cite{schneider-92, Ulmer95,Jaroszynski95,nakamura99} where the Airy function has been applied to lensing at the fold). 
Finally, the transmission factor near a fold in the TA framework is given by
\beq
  F_{\rm TA} =C\,  e^{\ii (w T_0-\frac{\pi}{4})}\,\rho_c^{-1/3} \, w^{1/6}
  \mathrm{Ai}
  \left[(w^2/\rho_c)^{1/3} \,\tilde{y}_2\right], 
\label{eq:TA_general}
\eeq
where $C=(2\pi/|T_{11}|)^{1/2}$ and $\rho_c=|T_{222}|/2$ are expressed through derivatives evaluated at the fold, and hence depend parametrically on the parameter $b$. For the symmetric configuration we obtain $C=\sqrt{\pi}$ and 
\beq
    \rho_c = 2\sqrt{2}\, q^{-3}\left(8b^2-q\right)(q-2b^2)^{1/2}
\eeq
with $q(b)=(1+8b^2)^{1/2}-1$.

It is important to note that Eq.~\eqref{eq:TA_general} is specifically derived for the fold catastrophe and cannot be directly applied to other caustic types. However, the methodology used to obtain this result can be extended to higher-order catastrophes, such as the cusp, by mapping the time-delay function to the corresponding canonical diffraction integrals---standard forms in catastrophe theory (Airy for fold, Pearcey for cusp, etc.), which dominate the wavefield near caustic singularities \cite{berry80iv}. In the present work, we will not pursue this generalization.

\section{Uniform Asymptotics}
\label{sec:uat}

The wave field obtained within the TA framework, expressed through the Airy function, remains valid only in the immediate vicinity of the fold singularity. Further away, the time delay function deviates from the cubic polynomial associated with the fold catastrophe [Eq.~\eqref{eq:fold_expansion}], leading to the breakdown of the TA approximation. 
 In the short-wavelength limit, the wave field can instead be described using geometric rays far from the caustics. Yet, the GO and TA solutions do not smoothly connect, as they rely on Taylor expansions around different stationary points. In the GO approximation, these points are determined by quadratic polynomials, leading to a diverging field at the fold. In contrast, the TA framework is based on a cubic polynomial with distinct extrema of the time delay function, yielding a finite solution at the fold. A more robust approach would be to construct an asymptotic solution that remains valid across the singularity while also being expressed in terms of the geometric ray parameters that describe the wave field far from it.
This conceptual limitation---where the TA does not inherently involve variables defined by the physical rays---can be overcome using the mathematical framework of {\em Uniform Approximation} (UA), initially introduced by Chester et al.~\cite{Chester1957} and later applied to wave theory by Kravtsov \cite{kravtsov64}, Ludwig \cite{ludwig65,Ludwig66}, and Berry \cite{berry-69}.

The core idea of UA is to represent the wave field solution both near and far from the fold as a linear superposition of the Airy function and its derivative, $c_1\,\mathrm{Ai} + c_2\,\mathrm{Ai'}$. These two terms encapsulate the leading-order contributions of the ray expansion, effectively accounting for the combined influence of all higher-order derivatives. 
At the same time, they form the minimal set of functions (just two) required for a seamless asymptotic matching with the two geometric rays associated with the two coalescing stationary points.
For the following derivations, it would be helpful to introduce the "fold function"
\beqa
    {\cal A}(u) &=&
    \frac{\kappa}{2\pi}\int_{-\infty}^\infty 
    \exp{[\ii \kappa \left(u s + s^3/3 \right)]} \,\dd s \nonumber \\
    &=& \kappa^{1/6}\sqrt{2\pi} \mathrm{Ai}(\kappa^{2/3}u),
\label{eq:Airy-fold-function}
\eeqa
which is essentially the Airy function but with the explicit large parameter $\kappa$.
The values of the indices 2/3 and 1/6 are associated with simple caustics of the fold type \cite{berry80iv}. 
With the parametrization $w=\kappa/\rho_c$, $\tilde{y}_2=\rho_c \,u$, the TA solution from Eq.~\eqref{eq:TA_general} reduces to the fold function \eqref{eq:Airy-fold-function} with a prefactor determined by $b$:
\beq
F_{\rm TA} = \frac{1}{\sqrt{\ii \rho_c \,|T_{11}|}} e^{\ii wT_0} {\cal A}(u).
\label{eq:TA_Asym}
\eeq
We also need the derivative of the fold function, $\dd {\cal A}/\dd u$, which is given by
\beq
{\cal A}'(u)= \kappa^{5/6}\sqrt{2\pi} \mathrm{Ai}'(\kappa^{2/3}u).
\label{eq:Airy-fold-derivative}
\eeq
We seek a UA solution in the form of
\beq
F_{\rm UA} = [C_1 {\cal A}(u) +  \frac{C_2}{\ii \kappa}{\cal A}'(u) ] \,
e^{\ii w \tau_0}
\label{eq:UA-to-be-found}
\eeq
which represents a ray expansion near the fold.
All parameters $C_1$, $C_2$, $u$ and $\tau_0$ are functions of the source location and need to be determined by matching with the GO solutions.
To determine their explicit form, we first substitute the asymptotic expressions of the fold function \eqref{eq:Airy-fold-function} and its derivative \eqref{eq:Airy-fold-derivative}, valid for $u<0$ and $|u|\gg1$, into Eq.~\eqref{eq:UA-to-be-found}. These expressions hold within the caustic region, where two stationary points are real-valued:
\begin{align}
    {\cal A}(u) &\simeq \sqrt{2} \,(-u)^{-1/4} 
    \cos \left[ \frac{2}{3} \kappa \, (-u)^{3/2} - \frac{\pi}{4} \right],
    \label{eq:asympA}\\
    {\cal A}'(u) &\simeq \sqrt{2} \,(-u)^{1/4} \,\kappa \,
    \sin \left[ \frac{2}{3} \kappa \, (-u)^{3/2} - \frac{\pi}{4} \right].
    \label{eq:asymptoticsInside}
\end{align}
This leads to the following expression for the transmission factor:
\begin{multline}  
F \simeq  
\frac{1}{\sqrt{2\ii}} e^{\ii \frac{2}{3} \kappa(-u)^{3/2}
+\ii w \tau_0} 
\left[C_1 (-u)^{-1/4} - C_2 (-u)^{1/4} \right] \\  
+ \sqrt{\frac{\ii}{2}} e^{-\ii \frac{2}{3} \kappa(-u)^{3/2}
+\ii w \tau_0} 
\left[C_1 (-u)^{-1/4} + C_2 (-u)^{1/4} \right]
\label{eq:UA_asymp}
\end{multline}
The explicit form of the parameters can be determined by matching Eq.~\eqref{eq:UA_asymp} to the GO solution for the two stationary points, which can be obtained from Eq.~\eqref{F_GO}:
\begin{equation}
    F_{\rm GO2}=|\mu_M|^{1/2}\, e^{\ii w T_M}-\ii\, |\mu_S|^{1/2}\, e^{\ii w T_S}.
    \label{eq:GO2}
\end{equation}
Matching both the phases and amplitudes, 
we obtain a set of equations that lead to 
$\tau_0=(T_S+T_M)/2$ for the global time delay and
\beq
u=-\left[\frac{3}{4\rho_c}\,(T_S-T_M)\right]^{2/3}
\label{eq:u_match}
\eeq
for the argument of the fold function. The remaining constants are given by
\begin{align*}
    C_1 &= \frac{1}{\sqrt{2\ii}}\,(-u)^{1/4}\left[{|\mu_M|}^{1/2}+{|\mu_S|}^{1/2}\right],\\
    C_2 &= \frac{1}{\sqrt{2\ii}}\,(-u)^{-1/4}\left[{|\mu_M|}^{1/2}-{|\mu_S|}^{1/2}\right].
\end{align*}
Introducing these definitions into Eq.~\eqref{eq:UA-to-be-found}, and defining $\tau \equiv (3/4)\,(T_S-T_M)$, the final uniform asymptotics is fully expressed through the GO parameters: $T_M$, $T_S$, $\mu_M$, and $\mu_S$, namely
\begin{multline}
    F_{\rm UA} = \sqrt{\pi} e^{\ii w\tau_0} \times \\
    \Big\{(w\tau)^{1/6} \frac{1}{\sqrt{\ii}}({|\mu_M|}^{1/2}+{|\mu_S|}^{1/2})\,
    \mathrm{Ai}[-(w\tau)^{2/3}]  \\ 
      +\, (w\tau)^{-1/6}\sqrt{\ii}\,({|\mu_S|}^{1/2}-{|\mu_M|}^{1/2})\, \mathrm{Ai'}[-(w\tau)^{2/3}] \Big\} .
\end{multline}
At the fold, the arguments of $\mathrm{Ai}$ and $\mathrm{Ai'}$ are zero, as $\tau$ vanishes, while the GO amplitudes $\mu_j$ diverge. The prefactors of $\mathrm{Ai}$ and $\mathrm{Ai'}$ involve indeterminate forms of type $0 \times \infty$ and $\infty \times 0$, respectively, but explicit evaluation shows that both expressions remain finite. Thus, the apparent singularities cancel out.

Outside the caustic region (often referred to as the caustic shadow) for $u>0$, the two GO rays become complex, resulting in time delays that form a complex conjugate pair: $T_{\pm}=T_r\pm \ii T_i$. To ensure a physically meaningful solution, we discard the ray with a negative imaginary part, $T_i<0$, that grows as we move farther from the caustic.
The second ray, which has a positive imaginary part, $T_i>0$, leads to an asymptotically decaying solution away from the caustic and should match the asymptotic behavior of the fold function for positive arguments:
\beq
    {\cal A}(u) \simeq \frac{1}{\sqrt{2}}\, u^{-1/4}e^{-\frac{2}{3}\kappa \,u^{3/2}}
\eeq
for $u>0$ and $|u|\gg1$. Matching the UA solution 
$C_1^+ {\cal A}(u) \,e^{\ii w \tau_0^+}$ with the GO wave 
$\sqrt{\mu_+} e^{\ii w T_+}$, we get 
the transmission factor in the shadow region as
\begin{multline}
    F_{\rm UA}=2\sqrt{\pi}\, |\mu_+|^{1/2}(w\tau_+)^{1/6}\times \\ \mathrm{Ai} [(w\tau_+)^{2/3}]\, e^{\ii w \tau_0^+-\ii\phi^{+}},
\end{multline}
where $\tau_+=(3/2)\,\mathrm{Im}(T_+)$, 
$\tau_0^+=\mathrm{Re}(T_+)$, and $\phi^{+}=\mathrm{Arg}(\det \Hat{K}_+)/2$ represents the argument of the complex determinant of the Hessian matrix defined in Eq.~\eqref{eq:hessian}.

\section{Results}
\label{sec:results}

\subsection{Caustic width}

It is highly valuable to analyze the region near the caustic, where the wave amplitude is expected to be significantly magnified due to ray focusing. Key aspects of interest include defining the width of this zone, quantifying the amplification near the fold, determining the spacing between interference maxima, and understanding how these features depend on various parameters, such as the lens mass, wavelength (or frequency), and the separation between the masses in the binary system.

\begin{figure}[t]
    \centering
    \includegraphics[width=\columnwidth, trim=20 40 20 80, clip]{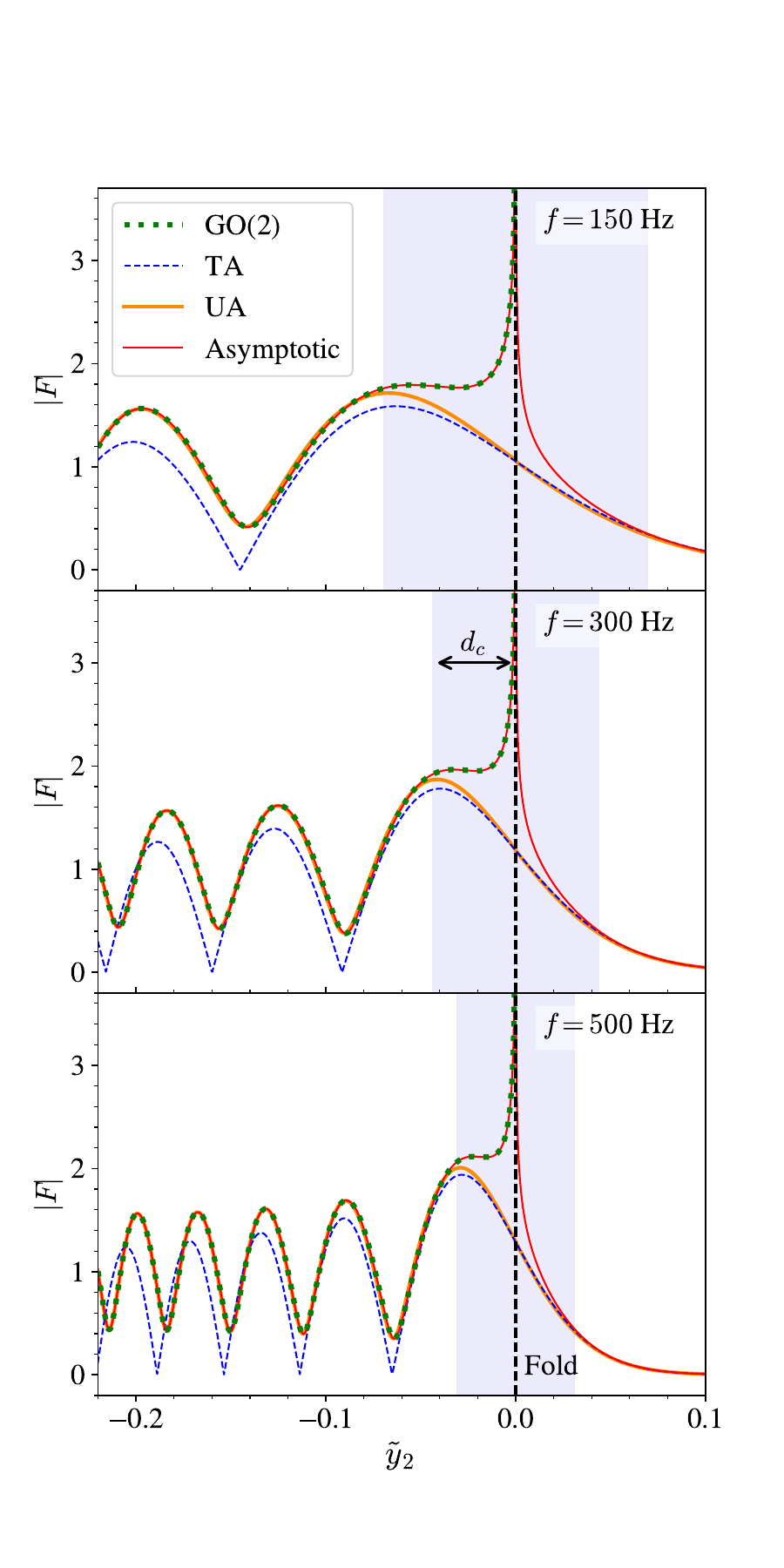}
    \caption{Absolute value of the transmission factor computed with the GO (green dotted line), TA (blue dashed line),  and UA (orange line) formulae. The UA results are also shown using the asymptotic expansions of the fold function and its derivative (red line). The binary separation is fixed to $b=0.7$ and the total redshifted mass of the lens to $M_{Lz}=5000\ \mathrm{M_\odot}$. 
    The black vertical dashed line indicates the position of the fold. The caustic zone $2d_c$ corresponds to the colored area.}
    \label{fig:UA-TA-GO-compare}
\end{figure}

Figure~\ref{fig:UA-TA-GO-compare} illustrates the behavior of the GO, UA, and TA approximations for the transmission factor
as a function of the source position along the $y_2$ axis.
The GO approximation diverges at the fold, whereas TA remains continuous. However, TA fails to match GO far from the fold. Near the fold, UA closely follows TA, ensuring continuity, whereas far from the fold, it aligns well with GO. This makes UA a robust approach, combining the accuracy of GO in the far region with the smooth transition of TA at the fold.
We focus on the behavior of the transmission factor within the frequency range detectable by the LVK network, which spans approximately 30 Hz to 1 kHz \cite{Aasi:2014jea,Acernese:2014hva,KAGRA:2020tym,Abbott2020}. In the figure, the frequencies used are 150 Hz, 300 Hz, and 500 Hz.
It would be useful to define the caustic region as the area where the GO approximation fails, while remaining approximately valid outside of it. As seen from the figure, this region can be characterized by the distance $d_c$ between the fold and the first interference maximum.
This distance can be determined by substituting $u=\tilde{y}_2/\rho_c$ into Eq.~\eqref{eq:u_match} and noting that the first maximum occurs when the phase difference between the GO rays is zero. Accounting for the Morse phase shift, this results in $w(T_S-T_M)=\pi/2$.
The resulting distance is given by
\beq
    d_c = (3\pi/8)^{2/3}\, \rho_c^{1/3}\, w^{-2/3}.
    \label{eq:caustic_width}
\eeq
Twice this distance, $2d_c$, can be referred to as the "caustic width" (see Fig.~\ref{fig:UA-TA-GO-compare}).
Recalling that we previously defined  $w=2\pi f t_M = 4\pi R_{\rm S}/\lambda$, 
we can conclude that the caustic width scales with the wavelength as $\sim\lambda^{2/3}$, and with the lens mass as $\sim M_{Lz}^{-2/3}$.
Based on this, we suggest the following expression for the caustic width (normalized to the Einstein radius):
\beqa
    2d_c &\simeq& 0.655\, \rho_c^{1/3}
    \left(\frac{\lambda}{2 R_{\rm S}}\right)^{2/3} \nonumber \\
    &\simeq& 0.2\, \rho_c^{1/3}
    \left(\frac{\lambda}{10^6 {\rm m}}\right)^{2/3}
    \left(\frac{M_{Lz}}{10^3 M_\odot}\right)^{-2/3}
    \label{eq:caustic_width_physical}
\eeqa
For $b=0.7$ (a half-distance between the masses is $0.7 R_{\rm E}$)  and 
a 300 Hz signal with a wavelength of $10^6$ m, we obtain $2d_c \approx 0.086 R_{\rm E}$ for $M = 5\times 10^3 M_\odot$, and $2d_c \sim 10^{-6} R_{\rm E}$ for $M = 10^{10} M_\odot$. Comparing these results with the lensing of electromagnetic waves, a wavelength in the visible spectrum of $0.5\,\mu$m results in $2d_c \sim 10^{-10} R_{\rm E}$ for $M = 5\times 10^3 M_\odot$, and $2d_c \sim 10^{-14} R_{\rm E}$ for $M = 10^{10} M_\odot$. These values illustrate that, for electromagnetic waves, GO provides an excellent approximation due to the small wavelength relative to the lensing scale. In contrast, for gravitational waves, diffraction effects become significant due to their longer wavelength, making GO inadequate over a broader region. The impact of wave effects depends on the lens mass: for a 300 Hz signal and a mass $M = 10^{10} M_\odot$, deviations from GO are significant only at distances to the caustic smaller than $10^{-6} R_{\rm E}$. In contrast, for lower masses, diffraction effects span a much larger region, emphasizing  the importance of the UA approximation in accurately modeling lensing of GWs.

We can now highlight several important conclusions: 
(i) On the one hand, the introduction of the caustic width allows us to estimate a new characteristic length scale $d_c$ associated with wave-optics effects. Due to the universality of the time-delay expansion near a fold, its dependence on the wavelength $\sim\lambda^{2/3}$ and the lens mass $\sim M_{Lz}^{-2/3}$ is also universal. Only the numerical coefficient in front is specific to the particular fold and the mass distribution of the lens.
(ii) On the other hand, as we will see in Sec.~\ref{sec:amplification}, the maximum amplification can be expressed in terms of the caustic width.
(iii) Finally, $d_c$ defines the region around the fold where the two overlapping rays can no longer be distinguished due to diffraction. That is, the rays can be physically distinguished only when $|\tilde{y}_2| > d_c$ within the interior zone.

In Appendix \ref{appendixB}, we employ the scaling in Eq.~\eqref{eq:caustic_width} to derive analytical expressions for the spacing between interference maxima.

\subsection{Amplification}
\label{sec:amplification}

To quantify the amplification near the fold, we use the asymptotics of the fold function \eqref{eq:Airy-fold-function}. As seen in Fig.~\ref{fig:UA-TA-GO-compare}, the asymptotic expressions of the Airy function and its derivative converge rapidly to the exact solution as we move away from the fold toward the interior region. This suggests that the first amplification peak, along with the subsequent oscillations, can be reliably estimated using the asymptotic expressions in Eqs.~\eqref{eq:asympA} and \eqref{eq:asymptoticsInside}.
Moreover, at the maximum of $\mathcal{A}(u)$—determined by the cosine—the derivative $\mathcal{A}'(u)$ vanishes, as it is governed by the sine. Therefore, the first amplification peak can be obtained by substituting the asymptotic form \eqref{eq:asympA} for $\mathcal{A}(u)$ into either Eq.~\eqref{eq:TA_Asym} or \eqref{eq:UA-to-be-found}.
The resulting Transitional Approximation at the first maximum yields:
\beq
|F_{\rm TA}|^* = \frac{1}{\sqrt{\rho_c \,|T_{11}|}} {\cal A}(u^*)
= \sqrt{2/|T_{11}|} (\rho_c\, d_c)^{-1/4},
\eeq
where ${\cal A}(u^*) = \sqrt{2} \,(-u^*)^{-1/4} $ is evaluated at $u^*$ given by $(-u^*)^{3/2}=(3\pi/8)\, w^{-1}\rho_c^{-1}$.
On the other hand, evaluating the Uniform Approximation at the maximum yields:
\beqa
&&|F_{\rm UA}|^* = C_1 {\cal A}(u^*)
= {|\mu_M|}^{1/2}+{|\mu_S|}^{1/2} \nonumber \\
&&\approx \sqrt{2/|T_{11}|} (\rho_c\, d_c)^{-1/4}\, \left[1+(1/6) (\rho_c')^2 d_c / \rho_c \right].
\eeqa
In the last line of this equation, the next-order term is included \cite{leontiev_II}.
Since the dominant term in both approximations agrees, it can be reliably used to estimate the peak amplification.
Given that GW detectors measure strain---sensitive to both amplitude and phase rather than intensity---the strain signal near the fold is amplified by a factor:
\beq
|F|_{\rm max} = F_0 \, d_c^{-1/4}  \quad
\text{with} \quad F_0 = \sqrt{2/|T_{11}|} \, \rho_c^{-1/4} 
\eeq
where the prefactor $F_0(b)$ depends only on the binary separation and is therefore fixed for a given lens mass distribution. In contrast, the final term, determined by the caustic width $d_c$, captures the frequency (or wavelength) dependence of the incoming wave.
For EM waves, where the observations are based on intensity, the relevant quantity is the magnification 
$\mu = |F|^2 = F_0^2 \, d_c^{-1/2}$.

\begin{figure}[t]
    \centering
    \includegraphics[width=\columnwidth]{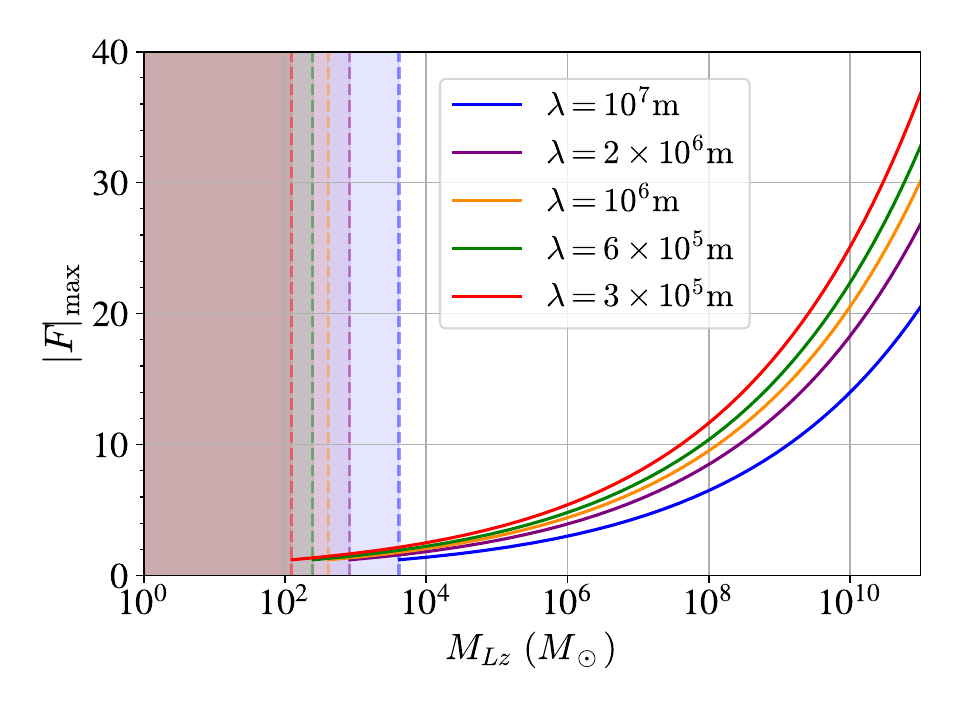}
     \caption{Maximum amplification near the fold, as given by Eq.~\eqref{eq:mu_max}, as a function of the redshifted lens mass. Results are shown for five different wavelengths, corresponding to the frequencies: 30 Hz, 150 Hz, 300 Hz, 500 Hz and 1000 Hz. Shadowed regions are included to indicate the threshold mass $M_{\rm min}$ \eqref{eq:M_min}.}
    \label{fig:max_ampli_vs_M}
\end{figure}

Therefore, the amplification $|F|_{\rm max}$ is governed by the characteristic scales of the caustic. Using Eq.~\eqref{eq:caustic_width_physical}, we find that it scales with the wavelength as $\sim\lambda^{-1/6}$, and with the lens mass as $\sim M_{Lz}^{1/6}$ as shown in Fig.~\ref{fig:max_ampli_vs_M}. This leads to the following expression:
\beq
|F|_{\rm max}
\simeq 2.5\ |T_{11}|^{-1/2}\, \rho_c^{-1/3}
    \left(\frac{\lambda}{10^6 {\rm m}}\right)^{-1/6}
    \left(\frac{M_{Lz}}{10^3 M_\odot}\right)^{1/6}
    \label{eq:mu_max}
\eeq

As an example, consider a binary lens system with $b = 0.7$ and a frequency of 500 Hz.
In this case, we find that the maximum amplification is 
$|F|_{\rm max} \simeq 2$ for a lens mass of $M = 5 \times 10^3 M_\odot$, and $|F|_{\rm max} \simeq 22.3$ for $M = 10^{10} M_\odot$. For comparison, the corresponding magnification values for electromagnetic (EM) waves will be the square of these values.

There is a limitation for Eq.\eqref{eq:mu_max} at low masses. Specifically, for smaller lens masses, the separation between the fold and its opposite counterpart may become comparable to or smaller than the fringe spacing, violating the condition for validity. 
From Eq.~\eqref{eq:TA_maxima}, the distance between the first and second interference maxima is given by $\tilde{y}_2(n=0)-\tilde{y}_2(n=1)\simeq2.15\ \rho_c^{1/3}\ w^{-2/3}$. This leads to the condition of validity: $2y_2^0\geq2.15\ \rho_c^{1/3}\ w^{-2/3}$. Expressing the dimensionless frequency $w$ in terms of physical parameters, the minimum lens mass is approximately
\beq
    M_{\rm min} \simeq 30  M_\odot \ ({y_2^0})^{-3/2}\rho_c^{1/2}\left(\frac{\lambda}{10^6\rm m}\right),
    \label{eq:M_min}
\eeq
which scales linearly with the wavelength.

\section{Conclusions}
\label{sec:conclusions}

When a GW passes near a massive lens, the full-wave transmission factor for arbitrary mass distributions in the lens plane can, in theory, be determined using the Kirchhoff diffraction integral. However, in practice, this approach becomes computationally expensive when the time delays between different paths—caused by the gravitational potential—are large compared to the wave period. In such cases, the integral becomes highly oscillatory and converges very slowly, making direct evaluation inefficient or even impractical. As a result, asymptotic methods are needed to obtain accurate and tractable approximations of the wave field in these regimes.

This study introduces a novel approach to understanding GW diffraction near caustics created by a binary lens system. 
The Uniform Approximation proves to be a robust method for modeling the GW field near the fold, addressing the limitations of both Geometrical Optics and Transitional Asymptotics methods. By ensuring continuity across the caustic, the UA effectively links wave behavior near the fold to the far-field GO description, providing a comprehensive solution for wave propagation in caustic regions.
While the TA [Eq.~(16)] and UA [Eqs.~(27), (29)] are based solely on the generic structure of fold caustics and are therefore broadly applicable, the specific numerical predictions we present are tailored to the binary lens model and may differ quantitatively for other lenses producing fold caustics.

We also introduce the concept of the caustic width, $d_c$, which defines a new length scale that characterizes the region where diffraction effects significantly alter wave propagation. The caustic width scales with the wavelength of the GW, $\sim \lambda^{2/3}$, and inversely with the lens mass, $\sim M_{Lz}^{-2/3}$. This universal scaling applies across various lens configurations and serves as a reliable tool for estimating diffraction regions in GW lensing studies. The caustic width $d_c$ marks the boundary within which overlapping rays become indistinguishable due to diffraction, suggesting that the number of detectable images for sources within this region may be reduced, with only rays outside the caustic remaining distinguishable.
Near the caustic, the maximum amplification—determined by the transition through the fold—can be reliably approximated using asymptotic expressions. This amplification is scaled with the caustic width as $~d_c^{-1/4}$.
The resulting amplification significantly enhances the wave signal, which could play a crucial role in detecting GWs near caustics.

Our findings suggest that GW detectors, such as those from the LIGO-Virgo-KAGRA network, may observe significant lensing effects, especially when the source is near caustics. 
Future studies should focus on extending the proposed methods to include a broader range of lens configurations. Additionally, investigating the impact of the caustic width and wave amplification on GW 
parameter estimation presents a promising direction for further research.

\section*{Acknowledgements}
\label{Acknowledgements}

The authors are grateful to Jose María Ezquiaga for insightful comments.
This work was supported by the Spanish Ministry of Science and Innovation through grants PID2021-125485NB-C22 and  CEX2019-000918-M funded by MCIN/AEI/10.13039/501100011033, as well as by the Agència de Gestió d’Ajuts Universitaris i de Recerca (AGAUR), Generalitat de Catalunya (grant SGR-2021-01069). 
AMS acknowledges financial support from the Institute of Cosmos Sciences of the University of Barcelona (ICCUB).

\appendix
\section{Images and caustics of a binary lens}
\label{appendixA}

The lens equation determines the source position $\mathbf{y}=(y_1,y_2)$ from which a ray passing through the lens plane at $\mathbf{x}=(x_1,x_2)$ originates. For a binary system, it takes the form:
\beq
    \mathbf{y}=\mathbf{x}-\frac{\gamma_1}{\left|\mathbf{x}-\mathbf{x_A}\right|^2}(\mathbf{x}-\mathbf{x_A})+\frac{\gamma_2}{\left|\mathbf{x}-\mathbf{x_B}\right|^2}(\mathbf{x}-\mathbf{x_B}),
    \label{lens_eq_vect}
\eeq
where $\mathbf{x_A}$ and $\mathbf{x_B}$ are the positions of the two lensing masses on the lens plane, and $\gamma_1$ and $\gamma_2$ are the mass ratios defined in Sec.~\ref{sec:microlens}. If the $x_1$ axis aligns with the line connecting the masses and the origin is at its midpoint, the mass coordinates satisfy $\mathbf{x_A}=-\mathbf{x_B}=\mathbf{x_L}=(b,0)$.

\begin{figure}[b]
    \centering
    \includegraphics[width=1.0\columnwidth, trim=0cm 20 0 0, clip]{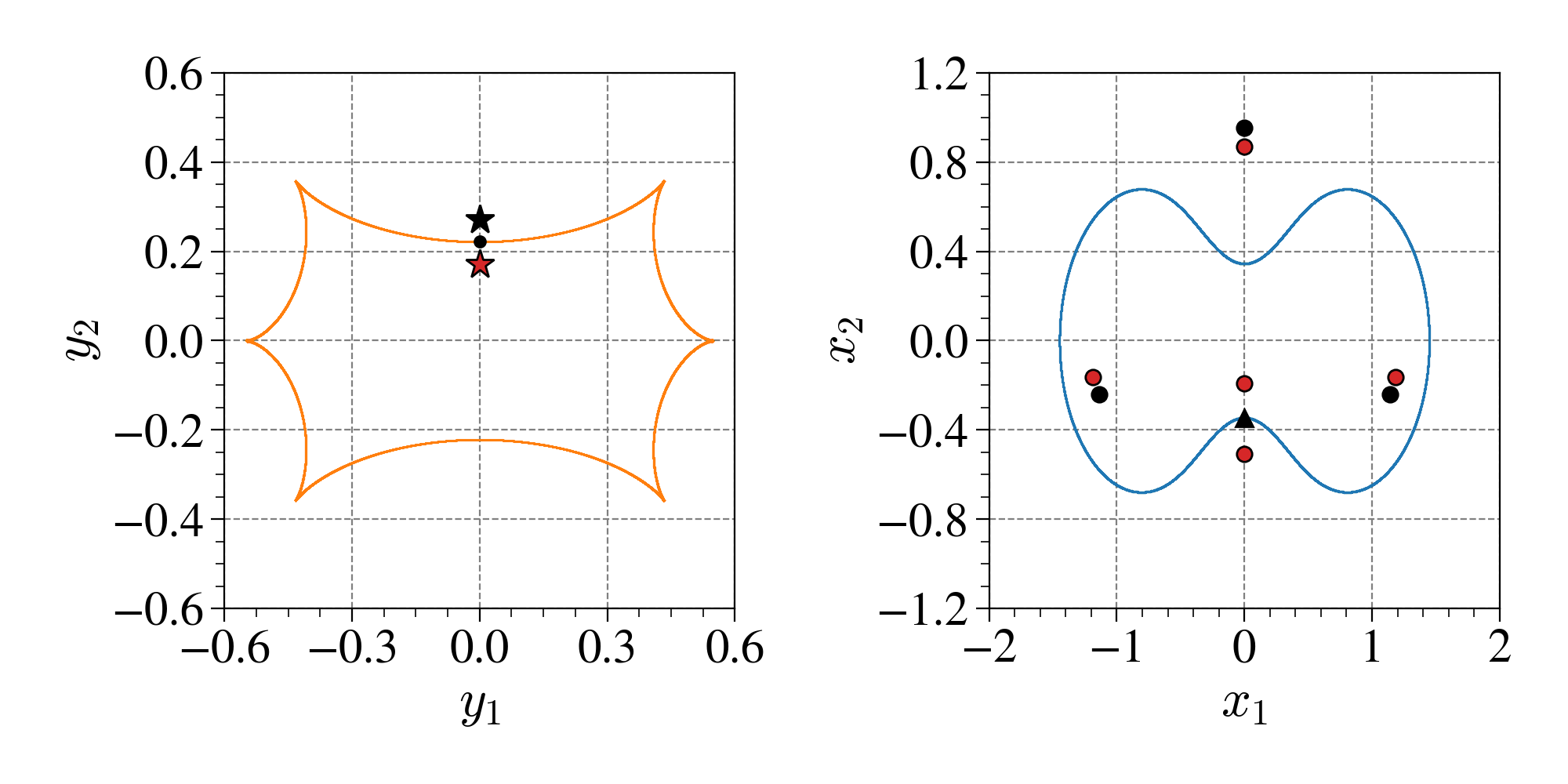}
    \caption{Source plane with caustics (left panel) and lens plane with critical lines (right panel) for a binary lens system with equal masses and $b = 0.7$. A source is placed at two different positions, both at a distance of $|\tilde{y}_2| = 0.05$ from the fold: one inside the caustic (red star) and one outside (black star). Their corresponding images in the lens plane are marked with dots of the same color. The location of the fold point is indicated by a black circle, while the point of coalescence of the two stationary points in the lens plane is denoted by a black triangle.}
    \label{fig:images_app}
\end{figure}
For computational convenience, we use the complex form of the lens equation, derived by representing vectors as complex numbers \cite{witt1990}. A vector $\mathbf{v}=(v_1,v_2)$ is written as $v=v_1+\ii v_2$, so the positions in the lens and source planes are $z=x_1+\ii x_2$ and $z_S=y_1+\ii y_2$, respectively, with $z_A$ and $z_B$ similarly representing the lensing masses. The lens equation then becomes:
\begin{equation}
    z_S=z-\frac{\gamma_1}{z^*-z_A^*}-\frac{\gamma_2}{z^*-z_B^*},
    \label{complex_lens}
\end{equation}
where $z^*$ is the complex conjugate of $z$. This equation can be reduced to a fifth-order complex equation. Solving it reveals the multiple images of a source at $z_S$. Depending on the source's position relative to the caustics, there are three distinct cases: a source outside the caustics produces three real images—two inside the critical lines and one outside; a source on the caustic results in an additional image on the critical line; and a source inside the caustics generates five images (see Fig. \ref{fig:images_app}). 

\begin{figure}[h]
    \centering
    \includegraphics[width=1.0\columnwidth]{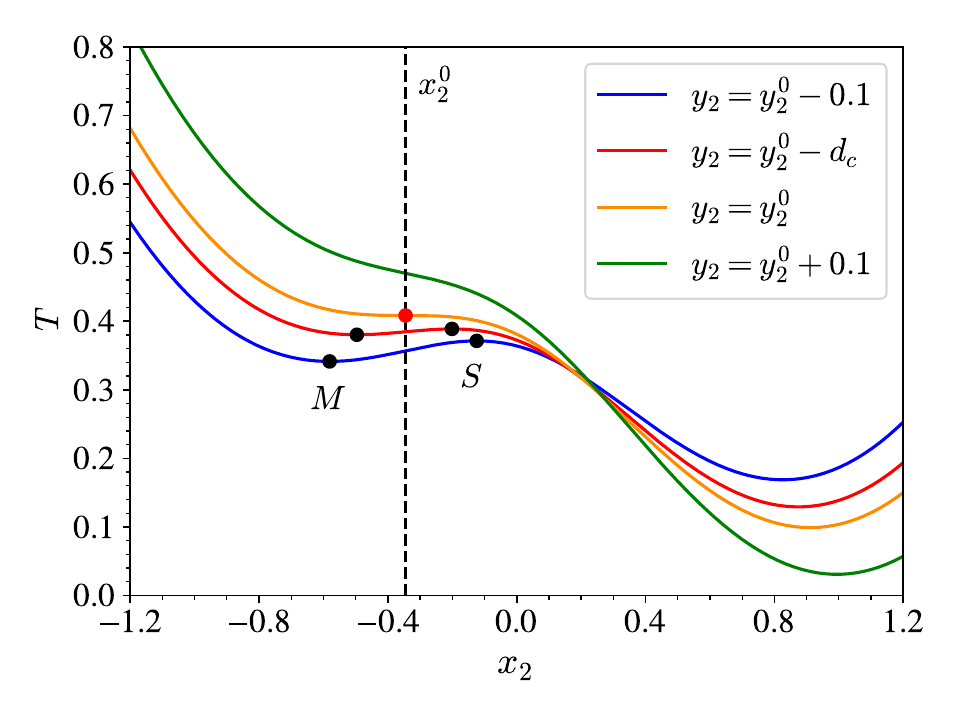}
    \caption{Time delay profile along the $x_2$-axis for various fixed source positions $y_2$.
    The black dots mark two stationary points---the minimum and the saddle point---which merge at $x_2 = x_2^0$ (vertical dashed line). The red dot represents the coalescence point corresponding to a source located exactly on the caustic. The caustic width $d_c$ is computed for a lens mass $M_{Lz} = 5\times10^3\ M_\odot$ and frequency $f = 300$ Hz.}
    \label{fig:time-delay_x2}
\end{figure}

To gain clearer insight into how additional images emerge as the source crosses the caustic, we show in Fig.~\ref{fig:time-delay_x2} the time delay function \eqref{eq:t_pre_expansion} along the $x_2$-axis for various source positions relative to the fold. When the source is inside the caustic region, the time delay exhibits two stationary points: a minimum (M) and a saddle point (S). As the source approaches the fold at $y_2^0$, the slope between the two stationary points gradually flattens (as shown by the blue and red lines). Upon reaching the caustic, the points merge on the critical line, causing the slope to vanish (orange line). Once the source crosses the caustic, these stationary points no longer exist in the real plane, evident by the absence of minima or maxima in the green line near $x_2^0$, and the corresponding images disappear.

\begin{figure*}
    \centering
    \includegraphics[width=\textwidth]{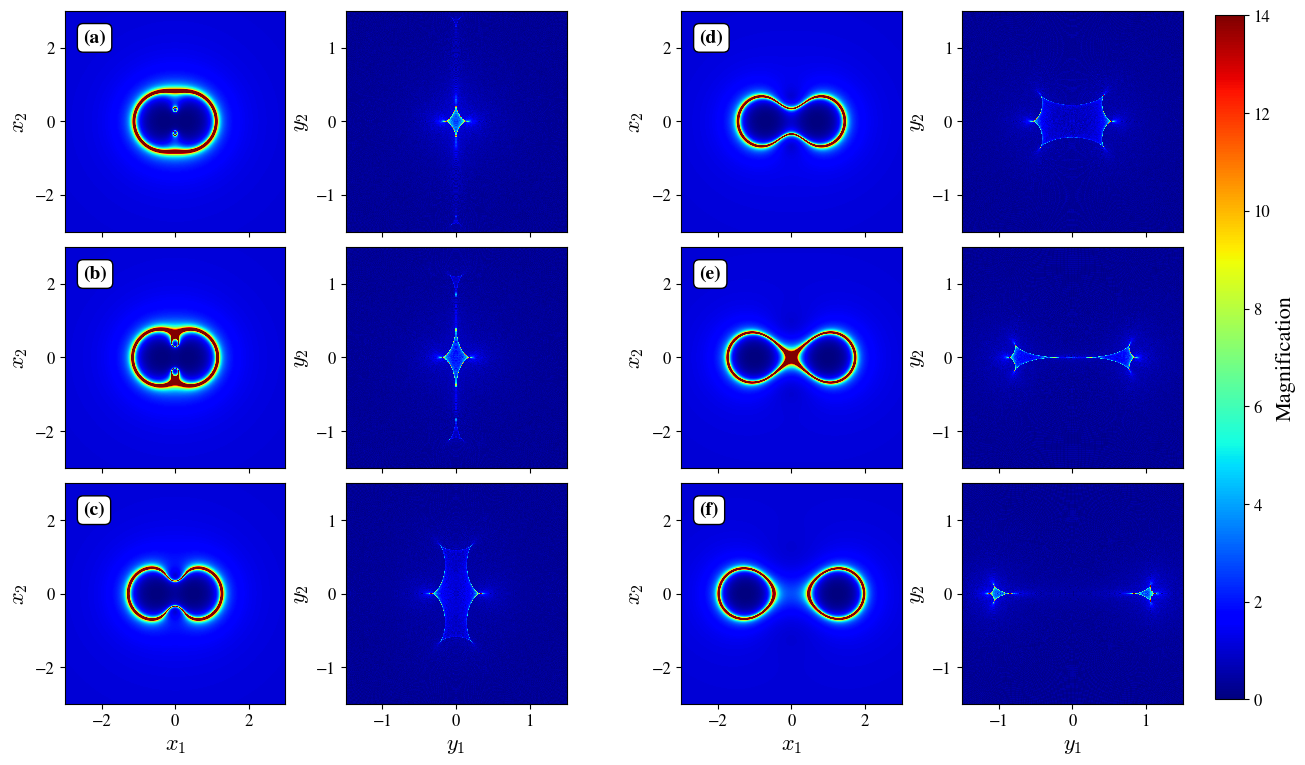}
    \caption{Magnification of a binary lens system with equal masses, 
    highlighting the morphological variations of the critical lines and caustics for different binary separations $b$: 0.3 (a), 0.35 (b), 0.5 (c), 0.7 (d), 1.0 (e), and 1.25 (f).
    For each subplot, the left panel displays the lens plane with critical lines, while the right panel shows the source plane with caustics. Magnification values were computed using Eq.~\eqref{eq:magnif} for the lens plane, and the inverse ray shooting method for the source plane \cite{tfg}.}
    \label{fig:magnification_maps}
\end{figure*}

The formation and evolution of images is intricately linked to the geometry of caustics, which, for a binary lens, is determined by the dimensionless distance between the masses, $2b$ \cite{schneider-weiss-86}, as shown in Fig.~\ref{fig:magnification_maps}. For $b > 1$, two separate extended caustics form. For $8^{-1/2} < b < 1$, a single caustic with six cusps is present. When $b < 8^{-1/2}$, three caustics appear: two triangular-shaped and one resembling an astroid with four cusps. The behavior of the caustics and critical lines varies as a function of the separation between the masses.

\begin{figure}
    \centering
    \includegraphics[width=\columnwidth]{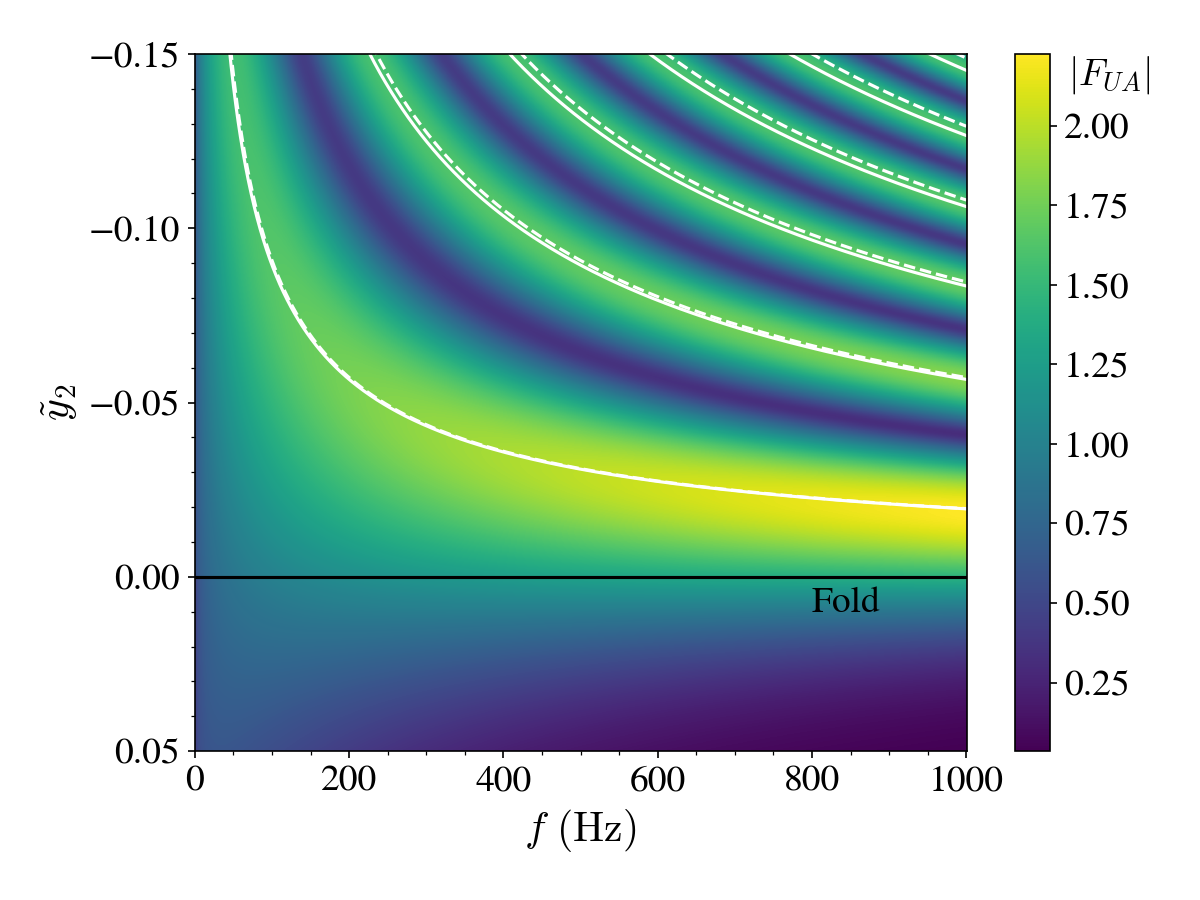}
    \caption{Absolute value of the transmission factor computed with UA as a function of the frequency and the separation of the source with respect to the fold. The binary separation is fixed at $b=0.7$, and the lens mass is set to $M_{Lz} = 5\times 10^{3} M\odot$. The fold location is indicated by a black line. The white continuous and dashed lines indicate the maxima computed with UA \eqref{eq:maxima_UA} and TA \eqref{eq:TA_maxima}, respectively.}
    \label{fig:amplification}
\end{figure}

\section{Interference Fringes}
\label{appendixB}

The maxima and minima of the transmission factor correspond to the positions of the source where constructive and destructive interference between the two GO rays occur, respectively. This condition is expressed as
\begin{equation}
\Delta \varphi=w(T_S-T_M)= 
\begin{cases} 
\pi/2+2\pi n, & \text{for maxima,} \\ 
3\pi/2+2\pi n, & \text{for minima,}
\end{cases}
\label{eq:max_min_TF_y2}
\end{equation}
where $\Delta \varphi$ is the phase difference between the two rays, and $n=0,1,2,...$. The additional term $\pi/2$ accounts for the Morse phase shift between the two stationary points. 

The time delay difference $T_S-T_M$ depends on the source's position relative to the caustic and appears in the UA approximation through the parameter $u$, as given by Eq.~\eqref{eq:u_match}. Near a caustic, $|\tilde{y}_2/T_{222}|\ll 1$, and $u$ can be expanded to second order in terms of the separation from the caustic, yielding: $u\simeq \tilde{y}_2/\rho_c-\left[3/2+(2\rho_c'/3)^2\right]\,\tilde{y}_2^2/(10\rho_c^2)$ \cite{leontiev_II}. Equating this expansion to Eq.~\eqref{eq:u_match}, and solving the quadratic equation for $\tilde{y}_2$, we obtain:
\begin{equation}
    \tilde{y}_2=D\left[1-\sqrt{1+2D^{-1}\left(3\rho_c^{1/2}\Delta\varphi/4w\right)}\right],
    \label{eq:maxima_UA}
\end{equation}
where $D=5\rho_c^{-1}\left[3/2+(2\rho_c'/3)^2\right]^{-1}$. By substituting the phase difference by Eq.~\eqref{eq:max_min_TF_y2}, we obtain the source positions corresponding to the extrema of the transmission factor. The dependence of these positions on the caustic curvature and the dimensionless frequency is nontrivial, resulting in a complex scaling for the fringe separation. Fig.~\ref{fig:amplification} shows a colormap of the absolute value of the transmission factor as a function of the source's distance from the fold and the frequency, where Eq.~\eqref{eq:maxima_UA} is used to visualize the maxima and minima.

Within the caustic zone, or in cases when the TA provides a good approximation to UA (i.e. large masses), the first order expansion suffices, resulting in: $u\simeq\tilde{y}_2/\rho_c$. Substituting this into Eq.~\eqref{eq:u_match} gives the source's distance to the fold as a function of the phase difference between the two GO rays:
\begin{equation}
    \tilde{y}_2=-\rho_c^{1/3}\,w^{-2/3}\,\left(3\Delta\varphi/4\right)^{2/3}.
    \label{eq:TA_maxima}
\end{equation}
This implies that in the frequency range 30 Hz - 1 kHz detectable by the LVK network, and for high masses, this solution effectively characterizes the interference fringe spacing near caustics, which scales with the lens mass as $\sim M_{Lz}^{-2/3}$, and with the wavelength as $\sim\lambda^{2/3}$, in agreement with catastrophe theory.

\bibliography{fold-caustic} 

\end{document}